\documentclass[twocolumn,showpacs,preprintnumbers,amsmath,amssymb]{revtex4}

\usepackage{graphicx}
\usepackage{dcolumn}
\usepackage{bm}
\usepackage{ulem}
\usepackage{amssymb}

\begin{document}
	
	
	\title{Pressure-induced magnetic transition exceeding 30 K in the Yb-based heavy fermion superconductor $\beta$-YbAlB$_4$}
	\author{Takahiro Tomita$^1$}
	\email{ttomita@issp.u-tokyo.ac.jp}
	\author{Kentaro Kuga$^1$}%
	\author{Yoshiya Uwatoko$^1$}%
	\author{Satoru Nakatsuji$^{1,2}$}
		\email{satoru@issp.u-tokyo.ac.jp}

	\affiliation{%
		$^1$Institute for Solid State Physics, The University of Tokyo, Kashiwanoha, Kashiwa, Chiba 277-8581, Japan}%
	\affiliation{%
			$^2$CREST, Japan Science and Technology Agency (JST), 4-1-8 Honcho Kawaguchi, Saitama 332-0012, Japan}%
	
	
	\begin{abstract}
		Measurements of the electric resistivity $\rho(T)$ under pressure up to 8~GPa were performed on high-quality single-crystals of the Yb-based heavy fermion system $\beta$-YbAlB$_4$ in the temperature range {\rm $2<T<$ 300~K}. In the resistivity data, we observed pressure-induced magnetic ordering above the critical pressure $P_{\rm c} \sim$ 2~GPa. {\rm Clear difference in the phase diagram under pressure using two types of pressure mediums indicates that the transition temperature may be further enhanced under application of uniaxial pressure.}  With pressure, this phase transition temperature $T_{\rm M}$ is enhanced reaching 32~K at 8~GPa, which is the highest transition temperature so far recorded for the Yb-based heavy fermion compounds. The power-law exponent $\alpha$ in $\rho=\rho_0+ AT^{\alpha}$ below $T_{\rm M}$ gradually changes from 3/2 to 5/2 with increasing pressure from 2 to 8~GPa. In contrast, the resistivity exhibits a $T$-linear behavior in the temperature range 2 $\le T \le$ 20~K and is insensitive to pressure below $P_{\rm c}$. In this pressure regime, the magnetization is also nearly independent of pressure and shows no anomaly above 2~K. Our results indicate that a quantum critical point for $\beta$-YbAlB$_4$ is also located near $P_{\rm c}$ in addition to {\rm the strange metal region near the ambient pressure}.
	\end{abstract}
	
	\maketitle

	\section{\label{sec:level1}Introduction}
	Intermetallic heavy-fermion (HF) compounds, mostly based on Ce, Yb or U, undergo a quantum phase transition at zero temperature induced by the competition between Kondo effects and intersite magnetic (RKKY) interaction.
	Various types of exotic phenomena such as non-Fermi-liquid behaviors, and unconventional superconductivity (SC) has been observed near a quantum critical point.
	The delicate balance has been tuned by varying control parameters such as magnetic field, doping, and pressure, as has been seen in the Ce-based quantum critical materials, e.g., CeCu$_2$(Si$_{1-x}$Ge$_{x}$)$_2$, \cite{Review1,Review2} CeIn$_3$, \cite{Review3}
	and the quasi-two-dimensional HF systems CeMIn$_5$ (M = Co, Rh, Ir). \cite{Review5,paglione-03}
	Significantly, the superconducting dome is found on the border of an antiferromagnetic (AF) phase in the pressure-temperature phase diagram. This suggests that, for these Ce-based HF compounds, the SC mechanism forming Cooper pairs is based on magnetic fluctuations.
	
	In contrast, the quantum-critical phenomena  in the Yb-based heavy fermion systems have not been explored as extensively as for Ce-based materials. A key subject would be on a possible novel quantum criticality expected based on  the electron--hole analogy of the 4$f$ state between 4$f^{1}$-Ce$^{3+}$ and 4$f^{13}$-Yb$^{3+}$.
	As a prototype of quantum-critical phenomena in the Yb-based materials, the non-Fermi liquid behavior in YbRh$_2$Si$_2$ has been well studied. {\rm \cite{Gegenwart2002,Friedemann2009}} This pure material exhibits weak antiferromagnetism below the N\'{e}el temperature $T_{\rm N}$ = 70~mK and can be driven to a HF quantum criticality by field-tuning at ambient pressure. Despite the worldwide research on novel phenomena in other HF materials, Yb-based HF superconductivity has never been observed, except in the recently discovered $\beta$-YbAlB$_4$ \cite{Review6}, neither at ambient conditions nor under hydrostatic pressure. 	{\rm Very recently, YbRh$_2$Si$_2$ has been also found to exhibit superconductivity at $T_{\rm c}$  = 1 mK in the antiferromagnetic ordered phase. \cite{Erwin}}
	
	The intermetallic compound $\beta$-YbAlB$_4$ is the first Yb-based HF superconductor with a transition temperature $T_{\rm c}$ = 80~mK and {\rm exhibits  zero-field quantum criticality (QC)}. \cite{Review7, Review6,Matsumoto2011Science}
	Hence, $\beta$-YbAlB$_4$ is a system highly appropriate for studying quantum critical phenomena at ambient pressure; i.e., the temperature dependence of the zero-field resistivity exhibits non-Fermi-liquid behavior, specifically quasi-linear $T$ dependence for 0.8~K $<T<$ 20~K and $T^{3/2}$ dependence for   $T_{\rm c}<T< 0.8$~K. \cite{Review6,Review7}
	Moreover, in the temperature range of $T_{\rm c} <T<$ 2~K, the system exhibits divergent susceptibility $\chi_{c} \propto T^{-1/2}$ under a low field of 50 mT applied along the $c$-axis and $-\ln T$ dependence of the magnetic part of the specific heat $C_{\rm M}/T$.  These anomalous thermodynamic behaviors suggest that the {\rm QC does not come from the  standard theory based on spin-density-wave-type instability}.
	In addition, strong valence fluctuations were observed in $\beta$-YbAlB$_4$ (Yb$^{\sim2.75+}$), in comparison with other Yb-based QC materials such as YbCu$_{5-x}$Al$_x$ (Yb$^{\sim2.95+}$), YbRh$_2$Si$_2$ (Yb$^{\sim2.9+}$), and YbNi$_3$Ga$_9$ (Yb$^{\sim2.9+}$). \cite{Review25,Yamamoto, Matsubayashi2015}
	The valence fluctuations possibly play an important role in understanding this unconventional quantum criticality in $\beta$-YbAlB$_4$ and thus exotic behavior may appear through controlling parameters such as external pressure, magnetic field, and chemical substitution.
	
	While maintaining sample quality, an external pressure affords efficient control to reveal the phase diagram near QCP in both Ce-based and Yb-based HF systems.
	In the Yb-based HF systems, long-range magnetic order is expected to be stabilized at high pressure. This is because the 4$f$ moments are generally known to become more localized by reducing volume, as observed for example in YbRh$_2$Si$_2$, YbCu$_2$Si$_2$, and YbNi$_{3}$Ga$_{9}$, \cite{Review10,Fernandez,Matsubayashi2015} in sharp contrast with the Ce-based counterparts. For $\beta$-YbAlB$_4$, of interest is how unconventional zero-field quantum criticality observed at ambient pressure is associated with magnetic order that is known to emerge under high pressure. \cite{Tomita}
	In this {\rm article}, we report the nature of the pressure-induced magnetic order using high-quality single crystals $\beta$-YbAlB$_4$ and discuss the quantum phase transition driven by pressure-tuning.
	
	\section{\label{sec:level2}Experimental Details}

		Single crystals of $\beta$-YbAlB$_4$ were grown by the aluminum self-flux method. \cite{Review12} To obtain high-quality single crystals, several crystals were selected using the residual resistivity ratio RRR $=\rho_{ab}(\mathrm{300 \  K})/\rho_{ab}(\mathrm{0  \ K})$. {\rm $\rho_{ab}(\mathrm{0 \  K})$ was estimated as  the linearly extrapolated value obtained at 0 K based on its temperature dependence observed within the 2-5 K range.}
		We spot-welded electrical contacts to the $ab$-plane surface of crystals using 20 $\mu$m\textit{$\phi$} Au wires.
		The temperature dependence of the electrical resistivity under various pressures up to 8~GPa was measured for the selected single crystals $\beta$-YbAlB$_4$ with RRR = 170 $\sim$ 250 (residual resistivity = 0.5 $\sim$ 1 $\mu \Omega$ cm) in the temperature region between 2 and 300~K using a cubic-anvil-type pressure cell {\rm \cite{Review13}} and a piston-type pressure cell. \cite{presscell} 
		In the cubic anvil pressure cell, the measurements were performed at a fixed pressure
		(where a constant load pressure is applied) and at variable temperatures in the 2 to 300~K range.
		Hydrostatic pressure up to 8~GPa was applied using two different pressure transmitting media, Daphne oil 7373 and  {\rm a mixture of Fluorinert 70 and Fluorinert 77}. DC magnetization measurements under pressure were performed using a commercial superconducting quantum interference device magnetometer (Quantum Design). \cite{Uedasquid} Load pressure was always changed at room temperature.

	\section{\label{sec:level3}Results and Discussion}
	\begin{figure}[h!]
		\includegraphics[scale =0.18]{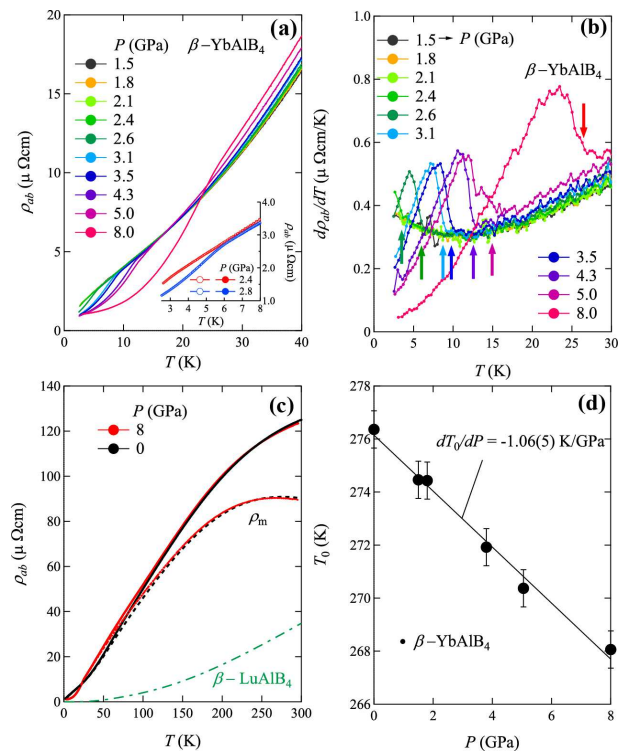}
		\caption{(Color online)  {\rm High pressure measurements of $\beta$-YbAlB$_4$ using Daphne oil.}  Pressure dependence of  (a) the in-plane resistivity $\rho_{ab}$ and (b) the temperature derivative of the resistivity $d \rho_{ab}/dT$ of $\beta$-YbAlB$_4$  (RRR = 200; red diamonds in Fig. \ref{fig-4})  up to pressure of 8~GPa. {\rm The inset shows $\rho_{ab}(T)$ in the warming (open circles) and cooling (closed circles) processes near the kink temperature at 2.4 and 2.8 GPa.}
			 Transition temperatures $T_{\rm{M}}$ are determined using $d\rho_{ab}(T)/dT$ values (see arrows).
			(c) Resistivity of $\beta$-YbAlB$_4$ under 0 (black symbol) and 8 (red) ~GPa and $\beta$-LuAlB$_4$ at ambient pressure over a wide temperature range from 2 to 300~K. The magnetic contribution $\rho_{\rm m}(T)$ is calculated by subtracting data of the ambient pressure $\beta$-LuAlB$_4$ from $\rho(T)$ of $\beta$-YbAlB$_4$.
			(d) Pressure dependence of the coherent peak at $T_0$ as determined by the maxima of $\rho_{\rm m}(T)$. See text for details.}
		\label{fig-1}
	\end{figure}
	
	Figure~\ref{fig-1}(a) displays the in-plane electrical resistivity $\rho_{ab}(T, P)$ of $\beta$-YbAlB$_4$ (RRR = 200) in the pressure range 0~$\le P \le$ 8~GPa measured using the cubic anvil pressure cell with the pressure medium Daphne 7373.
	While the temperature dependence of the resistivity $\rho_{ab}(T)$ is almost independent of pressure up to 2.1~GPa, $\rho_{ab}(T)$ exhibits a kink above $P=$ 2.4~GPa.
	The kink may well arise from a magnetic phase transition because the gradual drop in resistivity is normally associated with the loss of  scattering from disordered spins. Notably, it has been confirmed that Fe-substitution at the Al site  induces antiferromagnetic  ordering through the chemical pressure. \cite{Tomita, Review21} {\rm Moreover, a smooth evolution of $T_{\rm N}$ was observed as a function of pressure in $ \beta$-YbAlB$_4$, as well as  the Fe-substitution, and thus indicates that the pressure-induced phase in pure $\beta$-YbAlB$_4$ has the same type of AF order that was found in Fe-substituted $\beta$-YbAlB$_4$.}  {\rm Based on the fact that no hysteresis is observed around the transitions in the warming and cooling processes (inset of Fig.~\ref{fig-1} (a)), the transition is the 2nd order.}
	The kink temperature $T_{\rm M}$ gradually increases with increasing pressure and reaches   $\sim$ 30~K under 8~GPa. Here, we determined $T_{\rm M}$ using a kink found in the temperature derivative of $\rho_{ab}(T)$, $d \rho_{ab}(T)/dT$, [marked by arrows in Fig.~\ref{fig-1}(b)]. 
	The enhancement in $T_{\rm M}$ with pressure is expected in a magnetically ordered Yb Kondo lattice compound, as mentioned above, and differs from the decrease in $T_{\rm M}$ with pressure in Ce-based compounds.
	Figure~\ref{fig-1}(c) shows the temperature dependence of the resistivity $\rho_{ab}(T)$ and the magnetic contribution $\rho_{\rm m}(T)$ between 2 and 300~K for pressures of 0 and 8~GPa. The peak temperature  in $\rho_{\rm m}(T)$, $T_0$ [Fig.~\ref{fig-1}(d)] is generally  related to the coherence temperature $T_{\rm coh}$ for Kondo-lattice systems. \cite{Lavagna, Stewart}
	Thus, the coherence temperature is also suppressed with increasing pressure in $\beta$-YbAlB$_4$. {\rm Here, we also estimated $T_0$ by using the temperature dependence of  the absolute value of the resistivity that is calculated based on the actual length and cross-section of the sample under pressure, which are estimated from the X-ray diffraction measurements under pressure at room temperature. We found that this also gives the same $dT_0/dP$ slope within the error bar. \cite{Tomita}} This is a reasonable interpretation in the framework of the Doniach phase diagram. {\rm Here, the pressure suppresses the Kondo coupling scale and thus stabilizes the magnetic order, as indicated by the increase in $T_{\rm M}$ as a function of the pressure}.{\rm \cite{Doniach}}
	Moreover, the electrical resistivity at 300~K gradually increases with pressure, yielding a maximum value at 6~GPa and then decreases for $P<$ 8~GPa.
	
\begin{figure}[t!]
		
		\includegraphics[scale =0.22]{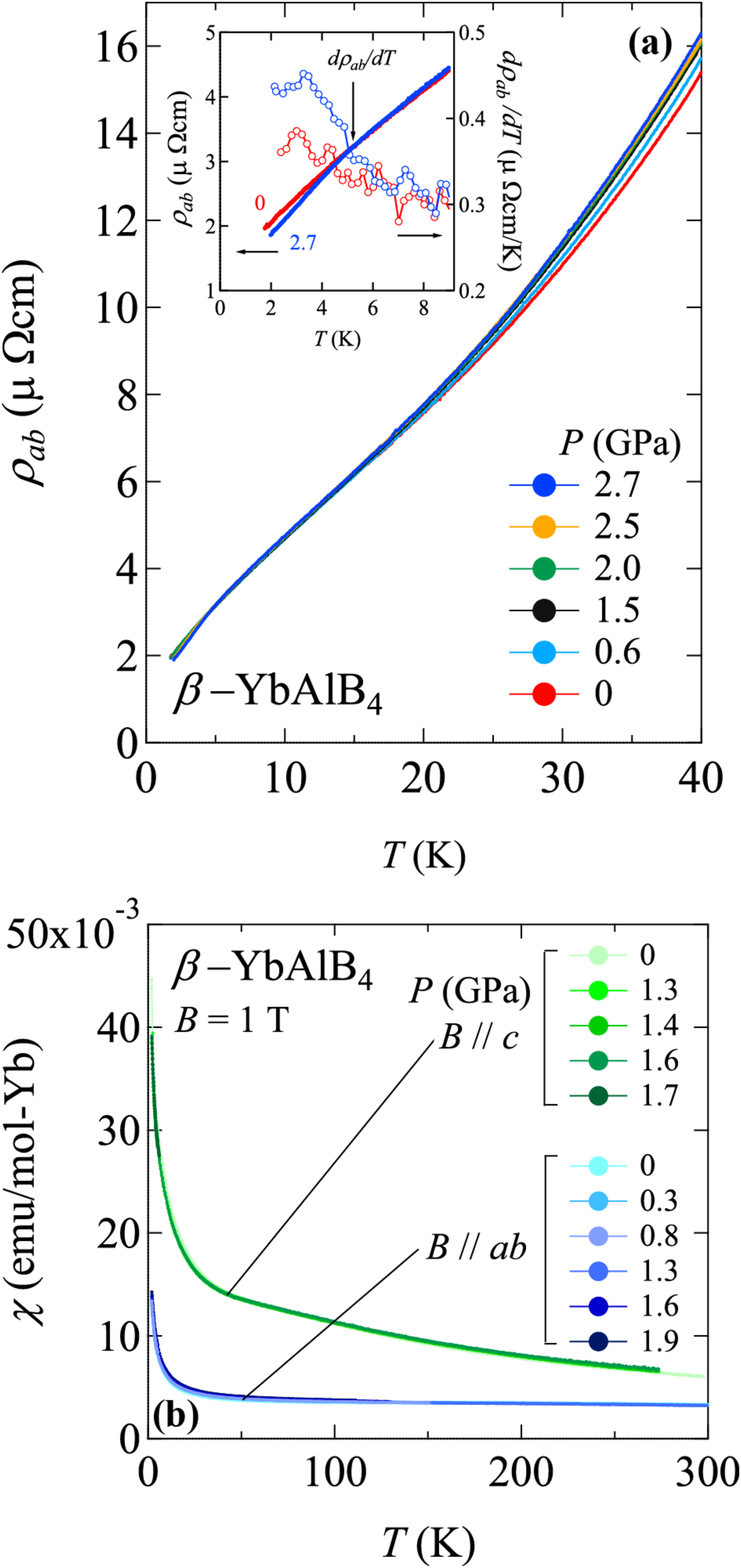}
		\caption{(Color online) {\rm Pressure measurements of $\beta$-YbAlB$_4$  using  Daphne oil}. (a) Temperature dependence of the electrical resistivity $\rho_{ab}(T)$ of $\beta$-YbAlB$_4$ (RRR = 100; {\rm indicated by red open circles in Fig. \ref{fig-4}}) for various pressures.
			Below 20~K, the exponent $\alpha$ defined by  $\rho_{ab}=\rho_o+AT^{\alpha}$ is nearly 1.
			The inset displays $\rho_{ab}(T)$ and $d\rho_{ab}(T)/dT$ for low temperatures; the arrow shows a pressure-induced transition temperature at $P$ = 2.7~GPa. (b) Temperature dependence of the magnetic susceptibility for both $B//ab$ and $B//c$ from 2 to 300~K under  various pressures. 
			See text for details.
		}
		\label{fig-2}
	\end{figure}

	Figure~\ref{fig-2}(a) shows the detailed pressure dependence of the electrical resistivity $\rho_{ab}(T, P)$ for $\beta$-YbAlB$_4$ (RRR = 100) in the pressure range 0 $\le P \le$ 2.7~GPa obtained using the piston-cylinder-type pressure cell.
	$\rho_{ab}(T)$ shows nearly $T$ linear dependence  below 20~K and remains unchanged for pressures up to $P=2$~GPa, which is consistent with the results below 2.1~GPa in Fig.~\ref{fig-1}(a).
	A change in the resistivity appears around 40~K corresponding to the peak found in the Hall coefficient $R_{\rm H}$, and may well come from the formation of the coherent state. \cite{EoinPRL2012} In particular, near this $T$ range close to 40~K, a systematic change in $\rho(T)$ with pressure was observed [Figs.~1(a) and 1(c)].
	As observed in the cubic anvil cell measurements under a pressure of 2.4~GPa, a kink is clearly seen near 4~K under $P\ge $ 2.4~GPa in both $\rho_{ab}(T)$ and the temperature derivative $d \rho_{ab}/dT$ (see inset of Fig.~\ref{fig-2}(a)). The same type of the transition behavior is also observed in Fig.~\ref{fig-1}(a).
	
	Consistent with the resistivity measurements below 2.1~GPa, the temperature-dependent Ising-like magnetization [Fig.~\ref{fig-2}(b)] shows no magnetic order and pressure independence for both $B//c$ and $B//ab$ in the pressure region 0$<P<$1.9~GPa and in the temperature region above 2~K.
	Hence, the magnitudes of both the Ising moment $I_{\rm z}$ and the Weiss temperature $\theta_{\rm W}$ {\rm are insensitive to} pressure below 2~GPa.\cite{matsumoto2011prb}
	The slight increases in magnetization below 20~K are larger than the previous observation \cite{Matsumoto2011Science} and are attributable to magnetic impurities on the sample surface. 
	 {\rm While we did try to clean the samples, our measurements suggest that the impurities were not fully removed from all the crystals used (30 pieces).}  Further measurements of the pressure dependence of the magnetization after washing the surface of impurities are necessary.
	
\begin{figure}[h!]
		\includegraphics[scale =0.22]{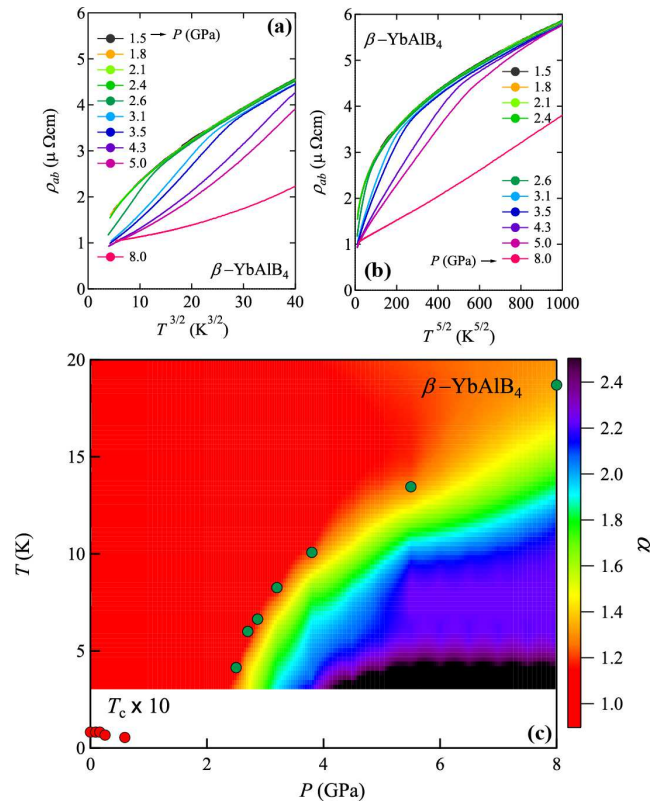}
		\caption{(Color online) {\rm High pressure measurements in $\beta$-YbAlB$_4$  (RRR=200; indicated by red diamonds in Fig. \ref{fig-4}) with Daphne oil.
			Low temperature in-plane resistivity vs (a) $T^{3/2}$ and (b) $T^{5/2}$ in $\beta$-YbAlB$_4$.
				(c) Pressure--temperature phase diagram for $\beta$-YbAlB$_4$ (RRR = 200; indicated by red circles in Fig. \ref{fig-4}) with a contour plot of the power law exponent $\alpha$, which is defined using $\rho=\rho_0+A T^{\alpha}$. $\alpha$ values between the measured pressures are interpolated linearly. }{\rm The red closed circles indicate the superconducting transition, $T_{\rm{c}} \times$ 10.} }
		\label{fig-3}
	\end{figure}

\begin{figure}[h!]
		\includegraphics[scale =0.22]{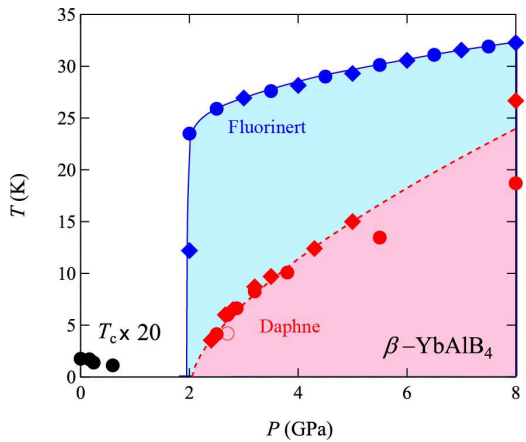}
		\caption{
			Pressure--temperature phase diagram for $\beta$-YbAlB$_4$ using two different pressure-transmitting media Daphne 7373 (red symbol) and Fluorinert (blue symbol), where data marked by the blue diamonds and circles are determined using single crystals with RRR = 250 and RRR = 170, respectively. Both red diamonds and circles indicate the results for {\rm two single crystals} with RRR = 200. Red diamonds and circles correspond to the results in Fig.~\ref{fig-1}(a) and in Fig.~\ref{fig-3}(c), respectively.
				The phase transition temperature $T_{\rm M}$ is determined from the anomalies in the temperature derivative of the resistivities, $d\rho_{ab}/dT$.
				Closed and open circles mark the transitions determined by the electrical resistance obtained from the cubic-anvil and piston-cylinder-type pressure measurements, respectively.
				The blue solid line is a guide for the eye.
				The red dashed line is a fit of our data to function $f \times (P-P_{\rm c0})^{2/3}$ expected for a 3D AF system. Here, $P_{\rm c0}=2.1$~GPa and $f=$ 7.3 K GPa$^{-2/3}$ are used. The superconducting transitions $T_{\rm c}$ (black closed circles, multiplied by a factor of 20) under pressure are also indicated. \cite{Tomita} See text for details.}
	\label{fig-4}
\end{figure}

	With regard to the temperature dependence of the resistivity above 2~K [Fig.~\ref{fig-1}(a)], the exponent $\alpha$ defined by $\rho_{ab}=\rho_0+ AT^{\alpha}$ gradually increases from 3/2 near 2~GPa to 5/2 above 3.1~GPa [{\rm Figs.~\ref{fig-3}(a) and ~\ref{fig-3}(b)}]; $\alpha=3/2$ is expected for the 3D AF-QCP based on the conventional spin fluctuation theory.	{\rm Figure~\ref{fig-3}(c) shows contour plots of the power law exponent	 $\alpha=\partial \ln(\rho-\rho_0)/\partial \ln T$ phase diagram of the other single-crystal of $\beta$-YbAlB$_4$ (RRR =200).  The wide $T$-linear region in $\rho_{ab}$ is observed in $ 0\le P \lesssim 2$ GPa at low temperatures. The exponent of $3/2$ appears  near the magnetic ordering temperature.}
	A similar power-law behavior, $\rho_{ab} \propto T^{3/2}$, for the resistivity drop appearing below the kink temperature has been observed in the pressure-tuned magnetically ordered state of MnSi near the quantum phase transition. \cite{Review17}
	In contrast, the $\alpha=5/2$ power law of the resistivity below $T_{\rm M}$ represents the electron--magnon scattering contribution \cite{Review18}, and has been seen in the AF phase of CeCu$_2$(Si$_{1-x}$Ge$_{x}$)$_2$. \cite{Review1}
	{\rm The exponent $\alpha$ changes smoothly from 3/2 to 5/2 with  increasing pressure.}
	The pressure dependence of $T_{\rm N}\propto(P-P_{\rm c0})^{2/3}$ as expected for a 3D AF system is fit to the data in Fig.~\ref{fig-4} using $P_{\rm c0}=$ 2.1~GPa. \cite{Hertz, Ueda, Mills} The fit is in good agreement with our data for Daphne 7373 below $\sim$ 4~GPa, where $\rho_{ab} \sim T^{3/2}$ is observed.
	All these results suggest that the phase transition is associated with an AF ordering. {\rm Thus, based on the $P-T$ phase shown in Fig.~\ref{fig-4}, the critical point is estimated to be $P_{\rm c0}=$ 2.1 GPa  near zero temperature.}
	
	\begin{figure}[h!]
		\includegraphics[scale =0.22]{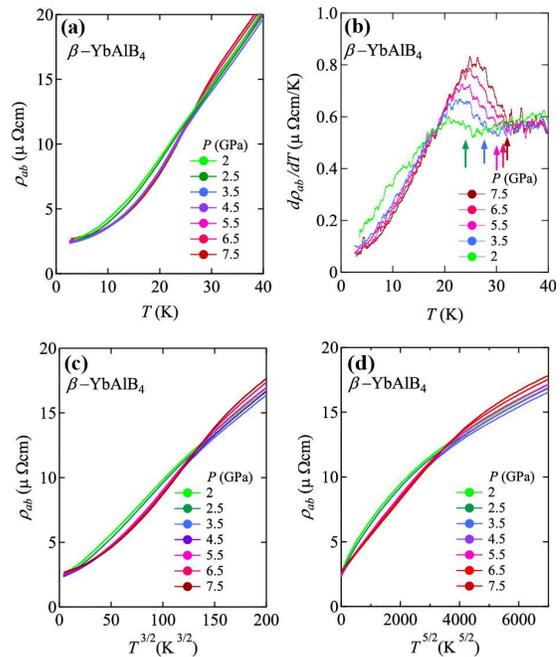}
		\caption{
			{\rm (Color online) High pressure measurements of $\beta$-YbAlB$_4$ (RRR = 200; indicated by blue circles in Fig. \ref{fig-4}) using  Fluorinert oil.  Temperature dependence of (a) the in-plane resistivity $\rho_{ab}$ and (b) the temperature derivative of the resistivity  $d\rho_{ab}(T)/dT$  of $\beta$-YbAlB$_4$  up to a pressure of 7.5~GPa. Transition temperatures $T_{\rm{M}}$ are determined using  $d\rho_{ab}(T)/dT$ values (see arrows). In-plane resistivity vs. (c) $T^{3/2}$ and (d) $T^{5/2}$ in $\beta$-YbAlB$_4$.}			
			 See text for details.}
		\label{fig-5}
	\end{figure}
		
	We further checked the resistivity under pressure up to 8~GPa using Fluorinert as pressure medium.{\rm  (see Figs. ~\ref{fig-5}(a) to \ref{fig-5}(d))} {\rm The $T_{\rm M}$ determined from the anomalies in $d\rho_{ab}/dT$ reaches at around 32 K under pressure of $P= 8 $ GPa. (See in Fig. \ref{fig-5}(b)) In the magnetically ordered phase observed above $P_{\rm c}$, the resistivity power-law exponent has 3/2 under pressure of 2 $\sim$ 3 GPa. (See in Fig. \ref{fig-5}(c)) }{\rm Under higher pressure, $T^{5/2}$ dependence emerges when the ordered temperature becomes as high as $\sim$ 20 K. (See in Fig. \ref{fig-5}(d)) } In Fig.~\ref{fig-4}, we summarize the pressure vs temperature phase diagram of $\beta$-YbAlB$_4$ for both pressure media, Daphene7373 (using the results in Fig.~\ref{fig-1}(a)) and Fluorinert.
	The two different phase diagrams indicate a large pressure-medium dependence. We measured the temperature and pressure dependence of the resistivity of different samples twice to check the phase transition using each pressure medium. Daphne oil is well-known to provide much better hydrostatic pressure than Fluorinert; {\rm in a cubic anvil cell, the applied pressure should be completely hydrostatic, while the pressure medium remains fluid. Under high pressure at room temperature, both the Fluorinert 70\&77 mixture and Daphne7373 oils becomes solidified at $P_{\rm s} \sim $  0.9 GPa and 2.2 GPa, respectively.  While the displacement of the six anvil tops was isotropic, the solidified medium causes a strongly anisotropic pressure distribution in the cell because the solidified pressure medium shows poor fluidity and the cell geometry is anisotropic. When $P>P_{\rm s}$, where $P$ is the current applied pressure and $P_{\rm s}$ is the solidified pressure, the deviation from the hydrostatic pressure, i.e., the so-called inhomogeneity, increases with lowering the solidified pressure and thus with the difference pressure $\Delta P$, which is defined as $\Delta P =P-P_{\rm s}$. The Fluorinert mixture is solidified at a lower pressure than the Daphne oil. The inhomogeneity for Fluorinert is thus higher than that for Daphne oil at higher pressures. \cite{hydrostatic, Yokogawa, Nakamura} Therefore, we}
	speculate that the difference in the phase diagram arises from the inhomogeneity in pressure that induces uni-axial stress in the crystal. A highly anisotropic bulk modulus is expected because the $ab$-plane network of the Boron covalent bond should be much harder than the $c$-axis ionic chain made of the Yb ionic bond. The anisotropic bulk modulus has been confirmed by the Fe-substitution dependence of the lattice parameter in $\beta$-YbAl$_{1-x}$Fe$_x$B$_4$, estimated from the X-ray measurements for Fe-substituted samples. \cite{Review21,Tomita}
	Moreover, a much sharper increase of the kink temperature was observed near $P_{\rm c}$ for the results taken using the pressure medium Fluorinert, suggesting the phase transition is of the first order.
	These indicate that the magnetically ordered phase is not connected with the SC phase and is separated from the quantum criticality observed for $B$ = 0 at ambient pressure. \cite{Matsumoto2011Science,Tomita}
	
	While almost pressure independent resistivity and magnetization are observed in the pressure range $0 \le P \le 2$~GPa, the magnetic phase transition of $\beta$-YbAlB$_4$ was suddenly found at $P>$ 2~GPa. Significantly, the phase transition of $\beta$-YbAlB$_4$ reaches {\rm 32~K} under 8~GPa. $T_{\rm M}$ is expected to be further enhanced at higher pressure. Such high magnetic transition temperatures over 10~K has never been achieved in Yb-based HF materials, e.g., YbInCu$_4$ ($T_{\mathrm{M}}$ = 3.3~K at $P$ = 7~GPa \cite{Review22}),
	YbCu$_2$Si$_2$ ($T_{\mathrm{M}}$ = {\rm 5~K at $P$ = 12~GPa} \cite{Fernandez}),
	YbRh$_2$Si$_2$ ($T_{\mathrm{M}}$ = {\rm 7~K at $P$ = 15~GPa }\cite{Dionicio}), and
	YbNi$_3$Ga$_9$ ($T_{\mathrm{M}}$ = 3~K at $P$ = 11~GPa \cite{Matsubayashi2015}).
	Hence, the observed transition temperature of  {\rm 32~K}  is the highest among the Yb-based HF compounds known so far.
	Generally, because the $f$-electron moment is sufficiently localized, the magnetic ordering temperatures in HF compounds is often observed below around 10~K. In particular, that of Yb-based HF should be lower than that of Ce-based HF because the Yb $f$-electron is more-localized compared with that for Ce.

	\section{\label{sec:level4}Conclusion}
	Using the electrical resistivity measurements obtained from high quality single crystals of $\beta$-YbAlB$_4$, we observed a pressure-induced magnetic transition above {\rm $P_{\rm c} =2.1$ ~GPa} ($P_{\rm c} \lesssim $ 2~GPa in Fluorinert). This phase transition appears at an extremely high temperature around  {\rm $\sim  30$~K} (32~K  under 8 GPa in Fluorinert). Furthermore, the dramatic difference in the phase diagram between those obtained by different pressure media indicates that the magnetic quantum phase transition may be weakly first order.
	In particular, both our measurements of resistivity and magnetization under pressure have revealed that the quantum critical phenomena are widely robust and no magnetic order is found down to 2~K in the intermediate pressure region 0 $<P<$ 2~GPa. This indicates that $\beta$-YbAlB$_4$ may have two quantum critical regions near ambient pressure and near $P_{\rm c}$, which are separated by the intermediate pressure region (Fermi liquid phase), suggesting a magnetic quantum critical point near {\rm $P_{\rm c} = 2.1$}~GPa. Indeed, this behavior has been actually confirmed by the low temperature measurements performed below 2 K. \cite{Tomita} 
	
	Generally, the HF compounds display quantum criticality at the border of magnetism. In the Ce-based HF superconductors, the superconducting phase near quantum criticality is connected to the AF phase. \cite{LonzarichNature} For $\beta$-YbAlB$_{4}$, the state showing unconventional critical behavior (the SC phase and ambient-pressure quantum criticality) is separated from this magnetically ordered state by the Fermi liquid phase. \cite{Tomita} Of high interest is the fact the non-Fermi liquid state at ambient pressure robustly persists up to a critical pressure of around 0.4 GPa as has been found in our former experiments. \cite{Tomita}
	These are not features that are expected for magnetically mediated superconductivity. The Yb-based HF with non-integer valence may exhibit critical phenomena with not only spin fluctuations but also valence fluctuations at quantum criticality. This valence fluctuation is presumably the key to understand  the origin of the extensive region of the quantum criticality found near ambient pressure and the extremely high ordering temperature of around  30~K found under high pressure. {\rm The pressure-induced ordering should be investigated in detail using magnetization and specific heat measurements under pressure.} The dramatic enhancement for $T_{\rm M}$ most likely caused by uni-axial pressure needs further study.

	\begin{acknowledgments}
		The authors thank K. Matsubayashi for help with the experiments and K. Ueda, K. Miyake, and S. Watanabe for fruitful discussions. The authors are also grateful to N. Horie for experimental assistance.
		This work was supported by Grants-in-Aid (Grant Nos. 25707030, 24740243 and 16H02209)  by CREST, Japan Science and Technology Agency, by Grants-in-Aid for Scientific Research on Innovative Areas (Grant Nos. 15H05882 and 15H05883) and Program for Advancing Strategic International Networks to Accelerate the Circulation of Talented Researchers (Grant No. R2604) from the Japanese Society for the Promotion of Science.
	\end{acknowledgments}
	
	\bibliography{iopart-num}

\begin{thebibliography}{38}
\expandafter\ifx\csname natexlab\endcsname\relax\def\natexlab#1{#1}\fi
\expandafter\ifx\csname bibnamefont\endcsname\relax
  \def\bibnamefont#1{#1}\fi
\expandafter\ifx\csname bibfnamefont\endcsname\relax
  \def\bibfnamefont#1{#1}\fi
\expandafter\ifx\csname citenamefont\endcsname\relax
  \def\citenamefont#1{#1}\fi
\expandafter\ifx\csname url\endcsname\relax
  \def\url#1{\texttt{#1}}\fi
\expandafter\ifx\csname urlprefix\endcsname\relax\def\urlprefix{URL }\fi
\providecommand{\bibinfo}[2]{#2}
\providecommand{\eprint}[2][]{\url{#2}}

\bibitem[{\citenamefont{Yuan et~al.}(2003)\citenamefont{Yuan, Grosche, Deppe,
  Geibel, Sparn, and Steglich}}]{Review1}
\bibinfo{author}{\bibfnamefont{H.~Q.} \bibnamefont{Yuan}},
  \bibinfo{author}{\bibfnamefont{F.~M.} \bibnamefont{Grosche}},
  \bibinfo{author}{\bibfnamefont{M.}~\bibnamefont{Deppe}},
  \bibinfo{author}{\bibfnamefont{C.}~\bibnamefont{Geibel}},
  \bibinfo{author}{\bibfnamefont{G.}~\bibnamefont{Sparn}}, \bibnamefont{and}
  \bibinfo{author}{\bibfnamefont{F.}~\bibnamefont{Steglich}},
  \bibinfo{journal}{Science} \textbf{\bibinfo{volume}{302}},
  \bibinfo{pages}{2104} (\bibinfo{year}{2003}).

\bibitem[{\citenamefont{Holmes et~al.}(2004)\citenamefont{Holmes, Jaccard, and
  Miyake}}]{Review2}
\bibinfo{author}{\bibfnamefont{A.~T.} \bibnamefont{Holmes}},
  \bibinfo{author}{\bibfnamefont{D.}~\bibnamefont{Jaccard}}, \bibnamefont{and}
  \bibinfo{author}{\bibfnamefont{K.}~\bibnamefont{Miyake}},
  \bibinfo{journal}{Phys. Rev. B} \textbf{\bibinfo{volume}{69}},
  \bibinfo{pages}{024508} (\bibinfo{year}{2004}).

\bibitem[{\citenamefont{Mathur et~al.}(1998)\citenamefont{Mathur, Grosche,
  Julian, Walker, Freye, Haselwimmer, and Lonzarich}}]{Review3}
\bibinfo{author}{\bibfnamefont{N.~D.} \bibnamefont{Mathur}},
  \bibinfo{author}{\bibfnamefont{F.~M.} \bibnamefont{Grosche}},
  \bibinfo{author}{\bibfnamefont{S.~R.} \bibnamefont{Julian}},
  \bibinfo{author}{\bibfnamefont{I.~R.} \bibnamefont{Walker}},
  \bibinfo{author}{\bibfnamefont{D.~M.} \bibnamefont{Freye}},
  \bibinfo{author}{\bibfnamefont{R.~K.~W.} \bibnamefont{Haselwimmer}},
  \bibnamefont{and} \bibinfo{author}{\bibfnamefont{G.~G.}
  \bibnamefont{Lonzarich}}, \bibinfo{journal}{Nature}
  \textbf{\bibinfo{volume}{394}}, \bibinfo{pages}{39} (\bibinfo{year}{1998}).

\bibitem[{\citenamefont{Tanator et~al.}(2007)\citenamefont{Tanator, Paglione,
  Petrovic, and Taillefer}}]{Review5}
\bibinfo{author}{\bibfnamefont{M.~A.} \bibnamefont{Tanator}},
  \bibinfo{author}{\bibfnamefont{J.}~\bibnamefont{Paglione}},
  \bibinfo{author}{\bibfnamefont{C.}~\bibnamefont{Petrovic}}, \bibnamefont{and}
  \bibinfo{author}{\bibfnamefont{L.}~\bibnamefont{Taillefer}},
  \bibinfo{journal}{Science} \textbf{\bibinfo{volume}{316}},
  \bibinfo{pages}{1320} (\bibinfo{year}{2007}).

\bibitem[{\citenamefont{Paglione et~al.}(2003)\citenamefont{Paglione, Tanatar,
  Hawthorn, Boaknin, Hill, Ronning, Sutherland, Taillefer, Petrovic, and
  Canfield}}]{paglione-03}
\bibinfo{author}{\bibfnamefont{J.}~\bibnamefont{Paglione}},
  \bibinfo{author}{\bibfnamefont{M.~A.} \bibnamefont{Tanatar}},
  \bibinfo{author}{\bibfnamefont{D.~G.} \bibnamefont{Hawthorn}},
  \bibinfo{author}{\bibfnamefont{E.}~\bibnamefont{Boaknin}},
  \bibinfo{author}{\bibfnamefont{R.~W.} \bibnamefont{Hill}},
  \bibinfo{author}{\bibfnamefont{F.}~\bibnamefont{Ronning}},
  \bibinfo{author}{\bibfnamefont{M.}~\bibnamefont{Sutherland}},
  \bibinfo{author}{\bibfnamefont{L.}~\bibnamefont{Taillefer}},
  \bibinfo{author}{\bibfnamefont{C.}~\bibnamefont{Petrovic}}, \bibnamefont{and}
  \bibinfo{author}{\bibfnamefont{P.~C.} \bibnamefont{Canfield}},
  \bibinfo{journal}{Phys. Rev. Lett.} \textbf{\bibinfo{volume}{91}},
  \bibinfo{pages}{246405} (\bibinfo{year}{2003}).

\bibitem[{\citenamefont{Gegenwart et~al.}(2002)\citenamefont{Gegenwart,
  Custers, Geibel, Neumaier, Tayama, Tenya, Trovarelli, and
  Steglich}}]{Gegenwart2002}
\bibinfo{author}{\bibfnamefont{P.}~\bibnamefont{Gegenwart}},
  \bibinfo{author}{\bibfnamefont{J.}~\bibnamefont{Custers}},
  \bibinfo{author}{\bibfnamefont{C.}~\bibnamefont{Geibel}},
  \bibinfo{author}{\bibfnamefont{K.}~\bibnamefont{Neumaier}},
  \bibinfo{author}{\bibfnamefont{T.}~\bibnamefont{Tayama}},
  \bibinfo{author}{\bibfnamefont{K.}~\bibnamefont{Tenya}},
  \bibinfo{author}{\bibfnamefont{O.}~\bibnamefont{Trovarelli}},
  \bibnamefont{and} \bibinfo{author}{\bibfnamefont{F.}~\bibnamefont{Steglich}},
  \bibinfo{journal}{Phys. Rev. Lett.} \textbf{\bibinfo{volume}{89}},
  \bibinfo{pages}{056402} (\bibinfo{year}{2002}).

\bibitem[{\citenamefont{Friedemann et~al.}(2009)\citenamefont{Friedemann,
  Westerkamp, Brando, Oeschler, S.Wirth, Gegenwart, Krellner, Geibel, and
  Steglich}}]{Friedemann2009}
\bibinfo{author}{\bibfnamefont{S.}~\bibnamefont{Friedemann}},
  \bibinfo{author}{\bibfnamefont{T.}~\bibnamefont{Westerkamp}},
  \bibinfo{author}{\bibfnamefont{M.}~\bibnamefont{Brando}},
  \bibinfo{author}{\bibfnamefont{N.}~\bibnamefont{Oeschler}},
  \bibinfo{author}{\bibnamefont{S.Wirth}},
  \bibinfo{author}{\bibfnamefont{P.}~\bibnamefont{Gegenwart}},
  \bibinfo{author}{\bibfnamefont{C.}~\bibnamefont{Krellner}},
  \bibinfo{author}{\bibfnamefont{C.}~\bibnamefont{Geibel}}, \bibnamefont{and}
  \bibinfo{author}{\bibfnamefont{F.}~\bibnamefont{Steglich}},
  \bibinfo{journal}{Nature Phys.} \textbf{\bibinfo{volume}{5}},
  \bibinfo{pages}{465} (\bibinfo{year}{2009}).

\bibitem[{\citenamefont{Nakatsuji et~al.}(2008)\citenamefont{Nakatsuji, Kuga,
  Machida, Tayama, Sakakibara, Karaki, Ishimoto, Yonezawa, Maeno, Pearson
  et~al.}}]{Review6}
\bibinfo{author}{\bibfnamefont{S.}~\bibnamefont{Nakatsuji}},
  \bibinfo{author}{\bibfnamefont{K.}~\bibnamefont{Kuga}},
  \bibinfo{author}{\bibfnamefont{Y.}~\bibnamefont{Machida}},
  \bibinfo{author}{\bibfnamefont{T.}~\bibnamefont{Tayama}},
  \bibinfo{author}{\bibfnamefont{T.}~\bibnamefont{Sakakibara}},
  \bibinfo{author}{\bibfnamefont{Y.}~\bibnamefont{Karaki}},
  \bibinfo{author}{\bibfnamefont{H.}~\bibnamefont{Ishimoto}},
  \bibinfo{author}{\bibfnamefont{S.}~\bibnamefont{Yonezawa}},
  \bibinfo{author}{\bibfnamefont{Y.}~\bibnamefont{Maeno}},
  \bibinfo{author}{\bibfnamefont{E.}~\bibnamefont{Pearson}},
  \bibnamefont{et~al.}, \bibinfo{journal}{Nature Phys.}
  \textbf{\bibinfo{volume}{4}}, \bibinfo{pages}{603} (\bibinfo{year}{2008}).

\bibitem[{\citenamefont{Schuberth et~al.}(2016)\citenamefont{Schuberth,
  Tippmann, Steinke, Lausberg, Steppke, Brando, Cornelius~Krellner, Yu, Si, and
  Steglich}}]{Erwin}
\bibinfo{author}{\bibfnamefont{E.}~\bibnamefont{Schuberth}},
  \bibinfo{author}{\bibfnamefont{M.}~\bibnamefont{Tippmann}},
  \bibinfo{author}{\bibfnamefont{L.}~\bibnamefont{Steinke}},
  \bibinfo{author}{\bibfnamefont{S.}~\bibnamefont{Lausberg}},
  \bibinfo{author}{\bibfnamefont{A.}~\bibnamefont{Steppke}},
  \bibinfo{author}{\bibfnamefont{M.}~\bibnamefont{Brando}},
  \bibinfo{author}{\bibfnamefont{C.~G.} \bibnamefont{Cornelius~Krellner}},
  \bibinfo{author}{\bibfnamefont{R.}~\bibnamefont{Yu}},
  \bibinfo{author}{\bibfnamefont{Q.}~\bibnamefont{Si}}, \bibnamefont{and}
  \bibinfo{author}{\bibfnamefont{F.}~\bibnamefont{Steglich}},
  \bibinfo{journal}{Science} \textbf{\bibinfo{volume}{351}},
  \bibinfo{pages}{485} (\bibinfo{year}{2016}).

\bibitem[{\citenamefont{Kuga et~al.}(2008)\citenamefont{Kuga, Karaki,
  Matsumoto, Machida, and Nakatsuji}}]{Review7}
\bibinfo{author}{\bibfnamefont{K.}~\bibnamefont{Kuga}},
  \bibinfo{author}{\bibfnamefont{Y.}~\bibnamefont{Karaki}},
  \bibinfo{author}{\bibfnamefont{Y.}~\bibnamefont{Matsumoto}},
  \bibinfo{author}{\bibfnamefont{Y.}~\bibnamefont{Machida}}, \bibnamefont{and}
  \bibinfo{author}{\bibfnamefont{S.}~\bibnamefont{Nakatsuji}},
  \bibinfo{journal}{Phys. Rev. Lett.} \textbf{\bibinfo{volume}{101}},
  \bibinfo{pages}{137004} (\bibinfo{year}{2008}).

\bibitem[{\citenamefont{Matsumoto
  et~al.}(2011{\natexlab{a}})\citenamefont{Matsumoto, Nakatsuji, Kuga, Karaki,
  Horie, Shimura, Sakakibara, Nevidomskyy, and Coleman}}]{Matsumoto2011Science}
\bibinfo{author}{\bibfnamefont{Y.}~\bibnamefont{Matsumoto}},
  \bibinfo{author}{\bibfnamefont{S.}~\bibnamefont{Nakatsuji}},
  \bibinfo{author}{\bibfnamefont{K.}~\bibnamefont{Kuga}},
  \bibinfo{author}{\bibfnamefont{Y.}~\bibnamefont{Karaki}},
  \bibinfo{author}{\bibfnamefont{N.}~\bibnamefont{Horie}},
  \bibinfo{author}{\bibfnamefont{Y.}~\bibnamefont{Shimura}},
  \bibinfo{author}{\bibfnamefont{T.}~\bibnamefont{Sakakibara}},
  \bibinfo{author}{\bibfnamefont{A.~H.} \bibnamefont{Nevidomskyy}},
  \bibnamefont{and} \bibinfo{author}{\bibfnamefont{P.}~\bibnamefont{Coleman}},
  \bibinfo{journal}{Science} \textbf{\bibinfo{volume}{331}},
  \bibinfo{pages}{316} (\bibinfo{year}{2011}{\natexlab{a}}).

\bibitem[{\citenamefont{Okawa et~al.}(2010)\citenamefont{Okawa, Matsunami,
  Ishizaka, Eguchi, Taguchi, Chainani, Takata, Yabashi, Tamasaku, Nishino
  et~al.}}]{Review25}
\bibinfo{author}{\bibfnamefont{M.}~\bibnamefont{Okawa}},
  \bibinfo{author}{\bibfnamefont{M.}~\bibnamefont{Matsunami}},
  \bibinfo{author}{\bibfnamefont{K.}~\bibnamefont{Ishizaka}},
  \bibinfo{author}{\bibfnamefont{R.}~\bibnamefont{Eguchi}},
  \bibinfo{author}{\bibfnamefont{M.}~\bibnamefont{Taguchi}},
  \bibinfo{author}{\bibfnamefont{A.}~\bibnamefont{Chainani}},
  \bibinfo{author}{\bibfnamefont{Y.}~\bibnamefont{Takata}},
  \bibinfo{author}{\bibfnamefont{M.}~\bibnamefont{Yabashi}},
  \bibinfo{author}{\bibfnamefont{K.}~\bibnamefont{Tamasaku}},
  \bibinfo{author}{\bibfnamefont{Y.}~\bibnamefont{Nishino}},
  \bibnamefont{et~al.}, \bibinfo{journal}{Phys.\ Rev.\ Lett.}
  \textbf{\bibinfo{volume}{104}}, \bibinfo{pages}{247201}
  (\bibinfo{year}{2010}).

\bibitem[{\citenamefont{Yamamoto et~al.}(2007)\citenamefont{Yamamoto, Yamaoka,
  Tsujii, Vlaicu, Oohashi, Sakakura, Tochio, Ito, Chainani, and
  Shin}}]{Yamamoto}
\bibinfo{author}{\bibfnamefont{K.}~\bibnamefont{Yamamoto}},
  \bibinfo{author}{\bibfnamefont{H.}~\bibnamefont{Yamaoka}},
  \bibinfo{author}{\bibfnamefont{N.}~\bibnamefont{Tsujii}},
  \bibinfo{author}{\bibfnamefont{A.~M.} \bibnamefont{Vlaicu}},
  \bibinfo{author}{\bibfnamefont{H.}~\bibnamefont{Oohashi}},
  \bibinfo{author}{\bibfnamefont{S.}~\bibnamefont{Sakakura}},
  \bibinfo{author}{\bibfnamefont{T.}~\bibnamefont{Tochio}},
  \bibinfo{author}{\bibfnamefont{Y.}~\bibnamefont{Ito}},
  \bibinfo{author}{\bibfnamefont{A.}~\bibnamefont{Chainani}}, \bibnamefont{and}
  \bibinfo{author}{\bibfnamefont{S.}~\bibnamefont{Shin}}, \bibinfo{journal}{J.
  Phys. Soc. Jpn.} \textbf{\bibinfo{volume}{76}}, \bibinfo{pages}{124705}
  (\bibinfo{year}{2007}).

\bibitem[{\citenamefont{Matsubayashi et~al.}(2015)\citenamefont{Matsubayashi,
  Hirayama, Yamashita, Ohara, Kawamura, Mizumaki, Ishimatsu, Watanabe,
  Kitagawa, and Uwatoko}}]{Matsubayashi2015}
\bibinfo{author}{\bibfnamefont{K.}~\bibnamefont{Matsubayashi}},
  \bibinfo{author}{\bibfnamefont{T.}~\bibnamefont{Hirayama}},
  \bibinfo{author}{\bibfnamefont{T.}~\bibnamefont{Yamashita}},
  \bibinfo{author}{\bibfnamefont{S.}~\bibnamefont{Ohara}},
  \bibinfo{author}{\bibfnamefont{N.}~\bibnamefont{Kawamura}},
  \bibinfo{author}{\bibfnamefont{M.}~\bibnamefont{Mizumaki}},
  \bibinfo{author}{\bibfnamefont{N.}~\bibnamefont{Ishimatsu}},
  \bibinfo{author}{\bibfnamefont{S.}~\bibnamefont{Watanabe}},
  \bibinfo{author}{\bibfnamefont{K.}~\bibnamefont{Kitagawa}}, \bibnamefont{and}
  \bibinfo{author}{\bibfnamefont{Y.}~\bibnamefont{Uwatoko}},
  \bibinfo{journal}{Phys. Rev. Lett.} \textbf{\bibinfo{volume}{114}},
  \bibinfo{pages}{086401} (\bibinfo{year}{2015}).

\bibitem[{\citenamefont{Knebel et~al.}(2006)\citenamefont{Knebel, Hassinger,
  Lapertot, Niklowitz, Sanchez, and Flouquet}}]{Review10}
\bibinfo{author}{\bibfnamefont{G.}~\bibnamefont{Knebel}},
  \bibinfo{author}{\bibfnamefont{E.}~\bibnamefont{Hassinger}},
  \bibinfo{author}{\bibfnamefont{G.}~\bibnamefont{Lapertot}},
  \bibinfo{author}{\bibfnamefont{P.~G.} \bibnamefont{Niklowitz}},
  \bibinfo{author}{\bibfnamefont{J.~P.} \bibnamefont{Sanchez}},
  \bibnamefont{and} \bibinfo{author}{\bibfnamefont{J.}~\bibnamefont{Flouquet}},
  \bibinfo{journal}{Physica B} \textbf{\bibinfo{volume}{378-380}},
  \bibinfo{pages}{68} (\bibinfo{year}{2006}).

\bibitem[{\citenamefont{{A. Fernandez-Pa\~{n}ella}
  et~al.}(2011)\citenamefont{{A. Fernandez-Pa\~{n}ella}, Braithwaite, Salce,
  Lapertot, and Flouquet}}]{Fernandez}
\bibinfo{author}{\bibnamefont{{A. Fernandez-Pa\~{n}ella}}},
  \bibinfo{author}{\bibfnamefont{D.}~\bibnamefont{Braithwaite}},
  \bibinfo{author}{\bibfnamefont{B.}~\bibnamefont{Salce}},
  \bibinfo{author}{\bibfnamefont{G.}~\bibnamefont{Lapertot}}, \bibnamefont{and}
  \bibinfo{author}{\bibfnamefont{J.}~\bibnamefont{Flouquet}},
  \bibinfo{journal}{Phys. Rev. B} \textbf{\bibinfo{volume}{84}},
  \bibinfo{pages}{134416} (\bibinfo{year}{2011}).

\bibitem[{\citenamefont{Tomita et~al.}(2015)\citenamefont{Tomita, Kuga,
  Uwatoko, Coleman, and Nakatsuji}}]{Tomita}
\bibinfo{author}{\bibfnamefont{T.}~\bibnamefont{Tomita}},
  \bibinfo{author}{\bibfnamefont{K.}~\bibnamefont{Kuga}},
  \bibinfo{author}{\bibfnamefont{Y.}~\bibnamefont{Uwatoko}},
  \bibinfo{author}{\bibfnamefont{P.}~\bibnamefont{Coleman}}, \bibnamefont{and}
  \bibinfo{author}{\bibfnamefont{S.}~\bibnamefont{Nakatsuji}},
  \bibinfo{journal}{Science} \textbf{\bibinfo{volume}{349}},
  \bibinfo{pages}{506} (\bibinfo{year}{2015}).

\bibitem[{\citenamefont{Macaluso et~al.}(2007)\citenamefont{Macaluso,
  Nakatsuji, Kuga, Thomas, Machida, Maeno, Fisk, and Chan}}]{Review12}
\bibinfo{author}{\bibfnamefont{R.~T.} \bibnamefont{Macaluso}},
  \bibinfo{author}{\bibfnamefont{S.}~\bibnamefont{Nakatsuji}},
  \bibinfo{author}{\bibfnamefont{K.}~\bibnamefont{Kuga}},
  \bibinfo{author}{\bibfnamefont{E.~L.} \bibnamefont{Thomas}},
  \bibinfo{author}{\bibfnamefont{Y.}~\bibnamefont{Machida}},
  \bibinfo{author}{\bibfnamefont{Y.}~\bibnamefont{Maeno}},
  \bibinfo{author}{\bibfnamefont{Z.}~\bibnamefont{Fisk}}, \bibnamefont{and}
  \bibinfo{author}{\bibfnamefont{J.~Y.} \bibnamefont{Chan}},
  \bibinfo{journal}{Chem. Mater.} \textbf{\bibinfo{volume}{19}},
  \bibinfo{pages}{1918} (\bibinfo{year}{2007}).

\bibitem[{\citenamefont{Mori et~al.}(2004)\citenamefont{Mori, Takahashi, and
  Takeshita}}]{Review13}
\bibinfo{author}{\bibfnamefont{N.}~\bibnamefont{Mori}},
  \bibinfo{author}{\bibfnamefont{H.}~\bibnamefont{Takahashi}},
  \bibnamefont{and}
  \bibinfo{author}{\bibfnamefont{N.}~\bibnamefont{Takeshita}},
  \bibinfo{journal}{High Press. Res.} \textbf{\bibinfo{volume}{24}},
  \bibinfo{pages}{225} (\bibinfo{year}{2004}).

\bibitem[{\citenamefont{Fujiwara et~al.}(2007)\citenamefont{Fujiwara,
  Matsumoto, Koyama-Nakazawa, Hisada, and Uwatoko}}]{presscell}
\bibinfo{author}{\bibfnamefont{N.}~\bibnamefont{Fujiwara}},
  \bibinfo{author}{\bibfnamefont{T.}~\bibnamefont{Matsumoto}},
  \bibinfo{author}{\bibfnamefont{K.}~\bibnamefont{Koyama-Nakazawa}},
  \bibinfo{author}{\bibfnamefont{A.}~\bibnamefont{Hisada}}, \bibnamefont{and}
  \bibinfo{author}{\bibfnamefont{Y.}~\bibnamefont{Uwatoko}},
  \bibinfo{journal}{Rev. Sci. Instrum.} \textbf{\bibinfo{volume}{78}},
  \bibinfo{pages}{073905} (\bibinfo{year}{2007}).

\bibitem[{\citenamefont{Ueda and Ueda}(2008)}]{Uedasquid}
\bibinfo{author}{\bibfnamefont{H.}~\bibnamefont{Ueda}} \bibnamefont{and}
  \bibinfo{author}{\bibfnamefont{Y.}~\bibnamefont{Ueda}},
  \bibinfo{journal}{Phys. Rev. B} \textbf{\bibinfo{volume}{77}},
  \bibinfo{pages}{224411} (\bibinfo{year}{2008}).

\bibitem[{\citenamefont{Kuga et~al.}(2012)\citenamefont{Kuga, Morrison,
  Treadwell, Chan, and Nakatsuji}}]{Review21}
\bibinfo{author}{\bibfnamefont{K.}~\bibnamefont{Kuga}},
  \bibinfo{author}{\bibfnamefont{G.}~\bibnamefont{Morrison}},
  \bibinfo{author}{\bibfnamefont{L.}~\bibnamefont{Treadwell}},
  \bibinfo{author}{\bibfnamefont{J.~Y.} \bibnamefont{Chan}}, \bibnamefont{and}
  \bibinfo{author}{\bibfnamefont{S.}~\bibnamefont{Nakatsuji}},
  \bibinfo{journal}{Phys. Rev. B} \textbf{\bibinfo{volume}{86}},
  \bibinfo{pages}{224413} (\bibinfo{year}{2012}).

\bibitem[{\citenamefont{Lavagna et~al.}(1983)\citenamefont{Lavagna, Lacroix,
  and Cyrot}}]{Lavagna}
\bibinfo{author}{\bibfnamefont{M.}~\bibnamefont{Lavagna}},
  \bibinfo{author}{\bibfnamefont{C.}~\bibnamefont{Lacroix}}, \bibnamefont{and}
  \bibinfo{author}{\bibfnamefont{M.}~\bibnamefont{Cyrot}}, \bibinfo{journal}{J.
  Phys. F: Met. Phys.} \textbf{\bibinfo{volume}{13}}, \bibinfo{pages}{1007}
  (\bibinfo{year}{1983}).

\bibitem[{\citenamefont{Stewart}(1984)}]{Stewart}
\bibinfo{author}{\bibfnamefont{G.~R.} \bibnamefont{Stewart}},
  \bibinfo{journal}{Rev. Mod. Phys.} \textbf{\bibinfo{volume}{56}},
  \bibinfo{pages}{755} (\bibinfo{year}{1984}).

\bibitem[{\citenamefont{Doniach}(1977)}]{Doniach}
\bibinfo{author}{\bibfnamefont{S.}~\bibnamefont{Doniach}},
  \bibinfo{journal}{Physica B+C} \textbf{\bibinfo{volume}{91}},
  \bibinfo{pages}{231} (\bibinfo{year}{1977}).

\bibitem[{\citenamefont{O'Farrell et~al.}(2012)\citenamefont{O'Farrell,
  Matsumoto, and Nakatsuji}}]{EoinPRL2012}
\bibinfo{author}{\bibfnamefont{E.~C.~T.} \bibnamefont{O'Farrell}},
  \bibinfo{author}{\bibfnamefont{Y.}~\bibnamefont{Matsumoto}},
  \bibnamefont{and}
  \bibinfo{author}{\bibfnamefont{S.}~\bibnamefont{Nakatsuji}},
  \bibinfo{journal}{Phys. Rev. Lett.} \textbf{\bibinfo{volume}{109}},
  \bibinfo{pages}{176405} (\bibinfo{year}{2012}).

\bibitem[{\citenamefont{Matsumoto
  et~al.}(2011{\natexlab{b}})\citenamefont{Matsumoto, Kuga, Tomita, Karaki, and
  Nakatsuji}}]{matsumoto2011prb}
\bibinfo{author}{\bibfnamefont{Y.}~\bibnamefont{Matsumoto}},
  \bibinfo{author}{\bibfnamefont{K.}~\bibnamefont{Kuga}},
  \bibinfo{author}{\bibfnamefont{T.}~\bibnamefont{Tomita}},
  \bibinfo{author}{\bibfnamefont{Y.}~\bibnamefont{Karaki}}, \bibnamefont{and}
  \bibinfo{author}{\bibfnamefont{S.}~\bibnamefont{Nakatsuji}},
  \bibinfo{journal}{Phys. Rev. B} \textbf{\bibinfo{volume}{84}},
  \bibinfo{pages}{125126} (\bibinfo{year}{2011}{\natexlab{b}}).

\bibitem[{\citenamefont{Pfleiderer et~al.}(2004)\citenamefont{Pfleiderer,
  Reznik, Pintschovius, v.~L\"{o}hneysen, Garst, and Rosch}}]{Review17}
\bibinfo{author}{\bibfnamefont{C.}~\bibnamefont{Pfleiderer}},
  \bibinfo{author}{\bibfnamefont{D.}~\bibnamefont{Reznik}},
  \bibinfo{author}{\bibfnamefont{L.}~\bibnamefont{Pintschovius}},
  \bibinfo{author}{\bibfnamefont{H.}~\bibnamefont{v.~L\"{o}hneysen}},
  \bibinfo{author}{\bibfnamefont{M.}~\bibnamefont{Garst}}, \bibnamefont{and}
  \bibinfo{author}{\bibfnamefont{A.}~\bibnamefont{Rosch}},
  \bibinfo{journal}{Nature} \textbf{\bibinfo{volume}{427}},
  \bibinfo{pages}{227} (\bibinfo{year}{2004}).

\bibitem[{\citenamefont{Antunes et~al.}(2006)\citenamefont{Antunes, Gospodinov,
  and Baibich}}]{Review18}
\bibinfo{author}{\bibfnamefont{A.~B.} \bibnamefont{Antunes}},
  \bibinfo{author}{\bibfnamefont{M.}~\bibnamefont{Gospodinov}},
  \bibnamefont{and} \bibinfo{author}{\bibfnamefont{M.~N.}
  \bibnamefont{Baibich}}, \bibinfo{journal}{Physica B}
  \textbf{\bibinfo{volume}{384}}, \bibinfo{pages}{47} (\bibinfo{year}{2006}).

\bibitem[{\citenamefont{Hertz}(1976)}]{Hertz}
\bibinfo{author}{\bibfnamefont{J.~A.} \bibnamefont{Hertz}},
  \bibinfo{journal}{Phys. Rev. B} \textbf{\bibinfo{volume}{14}},
  \bibinfo{pages}{1165} (\bibinfo{year}{1976}).

\bibitem[{\citenamefont{Moriya and Ueda}(2000)}]{Ueda}
\bibinfo{author}{\bibfnamefont{T.}~\bibnamefont{Moriya}} \bibnamefont{and}
  \bibinfo{author}{\bibfnamefont{K.}~\bibnamefont{Ueda}},
  \bibinfo{journal}{Advances in Physics} \textbf{\bibinfo{volume}{49}},
  \bibinfo{pages}{555} (\bibinfo{year}{2000}).

\bibitem[{\citenamefont{Millis}(1993)}]{Mills}
\bibinfo{author}{\bibfnamefont{A.~J.} \bibnamefont{Millis}},
  \bibinfo{journal}{Phys. Rev. B} \textbf{\bibinfo{volume}{48}},
  \bibinfo{pages}{7183} (\bibinfo{year}{1993}).

\bibitem[{\citenamefont{Hirayama et~al.}(2009)\citenamefont{Hirayama, Yamazaki,
  Fukazawa, Kohori, Takeshita, and Matsumoto}}]{hydrostatic}
\bibinfo{author}{\bibfnamefont{K.}~\bibnamefont{Hirayama}},
  \bibinfo{author}{\bibfnamefont{T.}~\bibnamefont{Yamazaki}},
  \bibinfo{author}{\bibfnamefont{H.}~\bibnamefont{Fukazawa}},
  \bibinfo{author}{\bibfnamefont{Y.}~\bibnamefont{Kohori}},
  \bibinfo{author}{\bibfnamefont{N.}~\bibnamefont{Takeshita}},
  \bibnamefont{and}
  \bibinfo{author}{\bibfnamefont{T.}~\bibnamefont{Matsumoto}},
  \bibinfo{journal}{J. Phys.: Conf. Ser.} \textbf{\bibinfo{volume}{150}},
  \bibinfo{pages}{012017} (\bibinfo{year}{2009}).

\bibitem[{\citenamefont{Yokogawa et~al.}(2007)\citenamefont{Yokogawa, Murata,
  Yoshino, and Aoyama}}]{Yokogawa}
\bibinfo{author}{\bibfnamefont{K.}~\bibnamefont{Yokogawa}},
  \bibinfo{author}{\bibfnamefont{K.}~\bibnamefont{Murata}},
  \bibinfo{author}{\bibfnamefont{H.}~\bibnamefont{Yoshino}}, \bibnamefont{and}
  \bibinfo{author}{\bibfnamefont{S.}~\bibnamefont{Aoyama}},
  \bibinfo{journal}{Jpn. J. Appl. Phys.} \textbf{\bibinfo{volume}{46}},
  \bibinfo{pages}{3636} (\bibinfo{year}{2007}).

\bibitem[{\citenamefont{Nakamura et~al.}(2000)\citenamefont{Nakamura, Goko,
  Hori, Uno, Kikugawa, and Fujita}}]{Nakamura}
\bibinfo{author}{\bibfnamefont{F.}~\bibnamefont{Nakamura}},
  \bibinfo{author}{\bibfnamefont{T.}~\bibnamefont{Goko}},
  \bibinfo{author}{\bibfnamefont{J.}~\bibnamefont{Hori}},
  \bibinfo{author}{\bibfnamefont{Y.}~\bibnamefont{Uno}},
  \bibinfo{author}{\bibfnamefont{N.}~\bibnamefont{Kikugawa}}, \bibnamefont{and}
  \bibinfo{author}{\bibfnamefont{T.}~\bibnamefont{Fujita}},
  \bibinfo{journal}{Phys. Rev. B} \textbf{\bibinfo{volume}{61}},
  \bibinfo{pages}{107} (\bibinfo{year}{2000}).

\bibitem[{\citenamefont{Kurita et~al.}(2006)\citenamefont{Kurita, Kano, Hedo,
  Uwatoko, Sarrao, Thompson, and Tozer}}]{Review22}
\bibinfo{author}{\bibfnamefont{N.}~\bibnamefont{Kurita}},
  \bibinfo{author}{\bibfnamefont{M.}~\bibnamefont{Kano}},
  \bibinfo{author}{\bibfnamefont{M.}~\bibnamefont{Hedo}},
  \bibinfo{author}{\bibfnamefont{Y.}~\bibnamefont{Uwatoko}},
  \bibinfo{author}{\bibfnamefont{J.~L.} \bibnamefont{Sarrao}},
  \bibinfo{author}{\bibfnamefont{J.~D.} \bibnamefont{Thompson}},
  \bibnamefont{and} \bibinfo{author}{\bibfnamefont{S.~W.} \bibnamefont{Tozer}},
  \bibinfo{journal}{Physica B} \textbf{\bibinfo{volume}{378-380}},
  \bibinfo{pages}{104} (\bibinfo{year}{2006}).

\bibitem[{\citenamefont{Dionicio et~al.}(2005)\citenamefont{Dionicio, Wilhelm,
  Sparn, Ferstl, Geibel, and Steglich}}]{Dionicio}
\bibinfo{author}{\bibfnamefont{G.}~\bibnamefont{Dionicio}},
  \bibinfo{author}{\bibfnamefont{H.}~\bibnamefont{Wilhelm}},
  \bibinfo{author}{\bibfnamefont{G.}~\bibnamefont{Sparn}},
  \bibinfo{author}{\bibfnamefont{J.}~\bibnamefont{Ferstl}},
  \bibinfo{author}{\bibfnamefont{C.}~\bibnamefont{Geibel}}, \bibnamefont{and}
  \bibinfo{author}{\bibfnamefont{F.}~\bibnamefont{Steglich}},
  \bibinfo{journal}{Physica B} \textbf{\bibinfo{volume}{359-361}},
  \bibinfo{pages}{50} (\bibinfo{year}{2005}).

\bibitem[{\citenamefont{Lonzarich}(2005)}]{LonzarichNature}
\bibinfo{author}{\bibfnamefont{G.~G.} \bibnamefont{Lonzarich}},
  \bibinfo{journal}{Nature Physics} \textbf{\bibinfo{volume}{1}},
  \bibinfo{pages}{11} (\bibinfo{year}{2005}).

\end{thebibliography}


\begin{thebibliography}{}
		
	\end{thebibliography}

\end{document}